\documentclass{svjour3}                     
\smartqed  
\usepackage{graphicx}
\journalname{jltp}
\begin{document}

\title{Inherent thermometry in a hybrid superconducting tunnel junction}

\author{H. Courtois \and Sukumar Rajauria \and P. Gandit \and F.W.J. Hekking \and B. Pannetier}

\institute{H. Courtois \at 
Low Temperature Laboratory, Helsinki University of Technology, P.O. Box 3500, 02015 TKK, Finland; Institut N\' eel, CNRS and Universit\'e Joseph Fourier, 25 Avenue des Martyrs, BP 166, 38042 Grenoble, France. \email{herve.courtois@grenoble.cnrs.fr}
\and Sukumar Rajauria \and P. Gandit  \and B. Pannetier \at
Institut N\' eel, CNRS and Universit\'e Joseph Fourier, 25 Avenue des Martyrs, BP 166, 38042 Grenoble, France.
\and F. W. J. Hekking \at
LPMMC, Universit\'e Joseph Fourier and CNRS, 25 Avenue des Martyrs, BP 166, 38042 Grenoble, France.}
\date{Received: date / Accepted: date}

\maketitle
\begin{abstract}
We discuss inherent thermometry in a Superconductor - Normal metal - Superconductor tunnel junction. In this configuration, the energy selectivity of single-particle tunneling can provide a significant electron cooling, depending on the bias voltage. The usual approach for measuring the electron temperature consists in using an additional pair of superconducting tunnel junctions as probes. In this paper, we discuss our experiment performed on a different design with no such thermometer. The quasi-equilibrium in the central metallic island is discussed in terms of a kinetic equation including injection and relaxation terms. We determine the electron temperature by comparing the micro-cooler experimental current-voltage characteristic with isothermal theoretical predictions. The limits of validity of this approach, due to the junctions asymmetry, the Andreev reflection or the presence of sub-gap states are discussed.
\keywords{Solid state cooling \and Superconducting tunnel junction \and Thermometry}
\PACS{74.50.+r \and 74.45.+c}
\end{abstract}

\section{Introduction}

A Normal metal (N) - Superconductor (S) tunnel junction, usually denoted as N-I-S where I stands for the Insulating barrier, is a very rich system which can be used for both electron cooling \cite{APL-Nahum2,APL-Pekola} or electron thermometry \cite{PRB-Rowell,APL-Nahum1} in the normal metal. The double junction geometry, i.e. S-I-N-I-S, brings the advantages of a double efficiency and a simplified fabrication. Combining electron cooling and thermometry on the same metallic island is usually made by having two pairs of superconducting tunnel probes connected to it. 

Let us consider the single quasi-particle tunneling current $I$ in a N-I-S  junction. It is given by:
\begin{equation}
I(V) = \frac{1}{eR_{N}}\int^{\infty}_{0} N_{S}(E)[f_{N}(E - eV) -f_{N}(E + eV)] dE,
\label{eq:symmetric-NIS}
\end{equation}
where $V$ is the voltage, $R_{N}$ is the normal state conductance, $f_{N}(E)$ is the electron distribution in the normal metal and
\begin{equation}
N_{S}(E) = \frac{|E|}{\sqrt{E^{2} - \Delta^{2}}}
\end{equation}
is the normalized BCS density of states in the superconductor, where $\Delta$ is the superconducting gap. We have used the anti-symmetry of the current-voltage characteristic to remove the superconductor energy distribution $f_S$ from the above expression. The tunnel current is thus insensitive to the temperature of the superconductor and depends solely on the electronic distribution in the normal metal and on the superconducting gap $\Delta$. 

\begin{figure}
\centering
\includegraphics[width=6 cm]{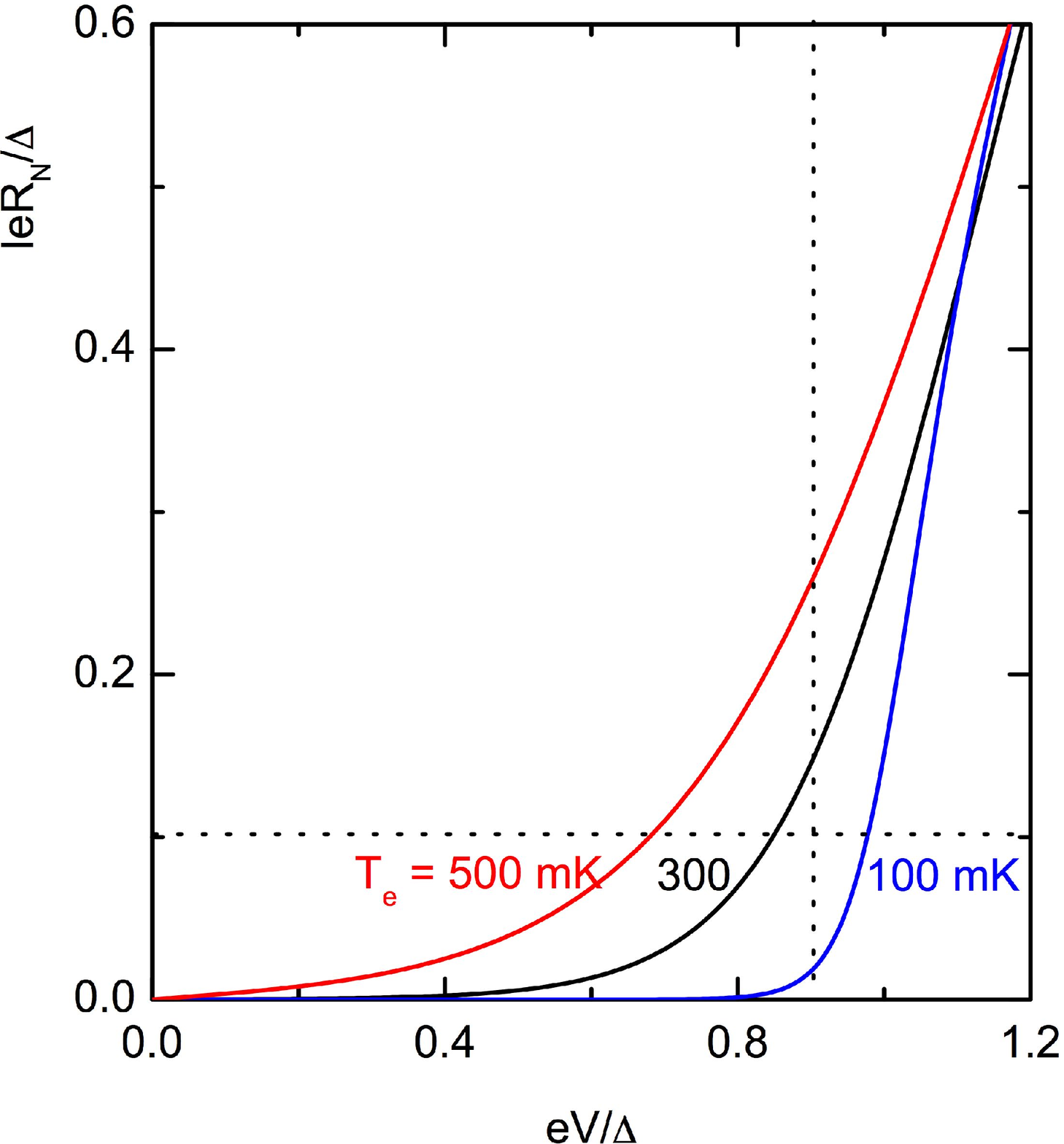}
\includegraphics[width=5.9 cm]{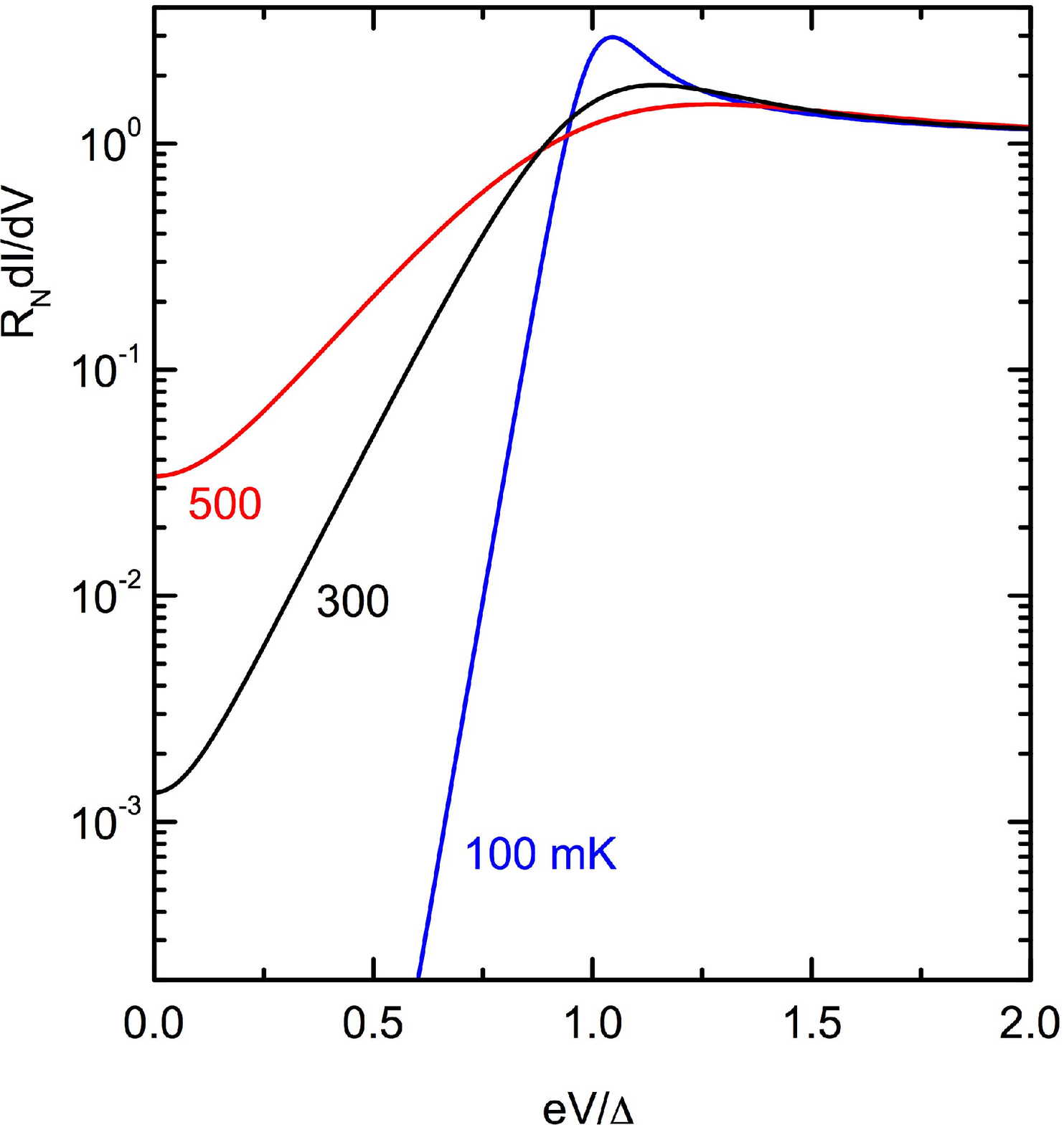}
\caption{(Color online) Calculated sub-gap current-voltage (left) and differential conductance (right) characteristics for a N-I-S junction at different temperatures. The superconducting gap is chosen as $\Delta$ = 0.225 meV, which is a typical value for Al, and the electronic  temperature T$_{e}$ is 100, 300 or 500 mK as indicated. On the left plot, the horizontal dotted line defines a current bias that can be used to measure the electronic temperatures through a voltage measurement. Similarly, a voltage bias (vertical dotted line) gives access to the temperature through a current measurement.}
\label{fig:isotherm}
\end{figure}
	
Fig. \ref{fig:isotherm} shows the calculated current-voltage and differential conductance characteristics for a N-I-S junction at thermal equilibrium at different electron temperatures T$_{e}$. The distribution function $f_N$ is then given by a Fermi function. For eV $> \Delta$, all the differential conductance curves merge at the level of the normal-state normalized conductance, equal to one. In the sub-gap regime, the current depends strongly on the temperature. For $k_BT_{e} \ll  \Delta$, the Fermi-Dirac function can be approximated by an exponential and the current through the N-I-S junction is then \cite{Solymar}:
\begin{equation}
I(V) \simeq I_{0}\exp[\frac{eV - \Delta}{k_BT_{e}}], \mbox{  with }I_{0} = \frac{\Delta}{eR_{n}}\sqrt{\frac{\pi k_BT_{e}}{2\Delta}}.
\end{equation}
This means that, in a logarithmic scale, the slope of the differential conductance $dI/dV$ versus the bias voltage $V$ plot is inversely proportional to the electron temperature. The horizontal dotted line on Fig. \ref{fig:isotherm} illustrates that when biasing the junction at a constant current, measuring the voltage gives directly the electronic temperature in the normal metal. Since R$_{n}$ and $\Delta$ are obtained from the experiment, there is no free fitting parameter involved. Voltage-biasing the junction and measuring the current provides the same type of information. 

\begin{figure}[t]
\centering
\includegraphics[width=0.27\linewidth]{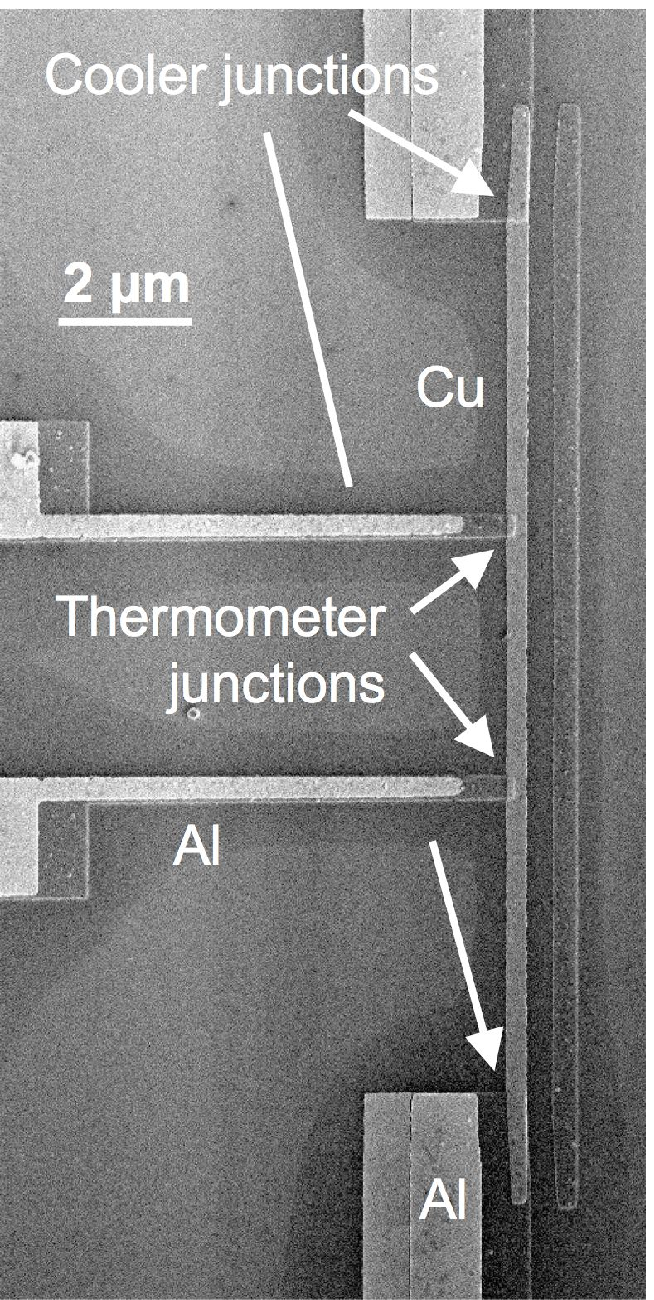}
\includegraphics[width=0.7\linewidth]{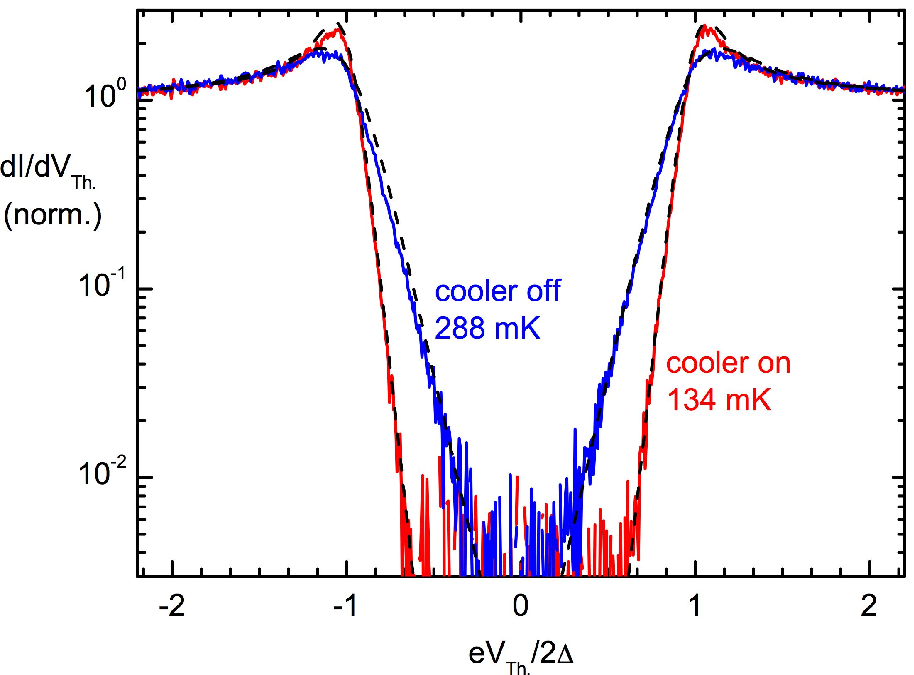}
\caption{(Color online) Left: Scanning electron micrograph of a micro-cooler with an external thermometer on the central normal metal island (Sample A). The cooler and the thermometer are made of two Al-AlO(x)-Cu junction in series. The area of thermometer junctions is 0.3 $\mu$m $\times$ 0.3 $\mu$m and that of cooler junctions is 1.5 $\mu$m $\times$ 0.3 $\mu$m. The central normal metal island is 14 $\mu$m long, 0.3 $\mu$m wide and 50 nm thick. Right: Sample A thermometer differential conductance versus voltage for two different cooler voltage biases. The state for zero cooler bias is called "cooler off". A cooler bias of 0.308 mV defines the "cooler on" state. The dashed lines are fits to the Eq. \ref{eq:symmetric-NIS}. The cryostat temperature is 275 mK and the gap is $\Delta$ = 0.46 mV.}
\label{SampleA-1}
\end{figure}

Fig. \ref{SampleA-1} left part shows the design of the first set of devices we have studied. The central normal metal (Cu) island is connected to two superconducting (Al) reservoirs through tunnel barriers of 0.3 $\times$ 1.2 $\mu$m$^2$ large area (top and bottom of Fig. \ref{SampleA-1} left part). This S-I-N-I-S junction is called the "cooler junction". The two 40 nm thick and 1.5 $\mu$m wide superconducting Al electrodes were in-situ oxidized in 0.2 mbar of oxygen for 3 min before the deposition of the central Cu island that is 14 $\mu$m long, 0.3 $\mu$m wide and 50 nm thick. The normal metal island has two additional small Al junctions on it (left of the image) with an area of 0.3 $\times$ 0.3 $\mu$m$^2$. These two junctions constitute a S-I-N-I-S thermometer.

Fig. \ref{SampleA-1} right part shows the thermometer differential conductance versus voltage at two different cooler voltage biases $V_{cooler}$. Both data are fitted with Eq. \ref{eq:symmetric-NIS} (dashed lines). In the "cooler off" state defined by $V_{cooler}$ = 0, the fitted normal metal electron temperature is 288 mK, which is close to the measured cryostat temperature. This agreement shows the good thermalization of the normal metal electrons to the cryostat temperature. In the "cooler on" state, the cooler junction is biased near the optimum voltage, V$_{cooler}$ = 0.308 mV. The same fitting procedure provides a normal metal electron temperature of 134 mK. This experiment illustrates that electron cooling can be accurately detected with a S-I-N-I-S thermometer. The quality of the fits is consistent with a thermal quasi-equilibrium of the electron population in the metal. Here, the electrons in the metallic island have cooled from the cryostat temperature of 288 mK down to 134 mK.

\begin{figure}[t]
\centering
\includegraphics[width=0.7\linewidth]{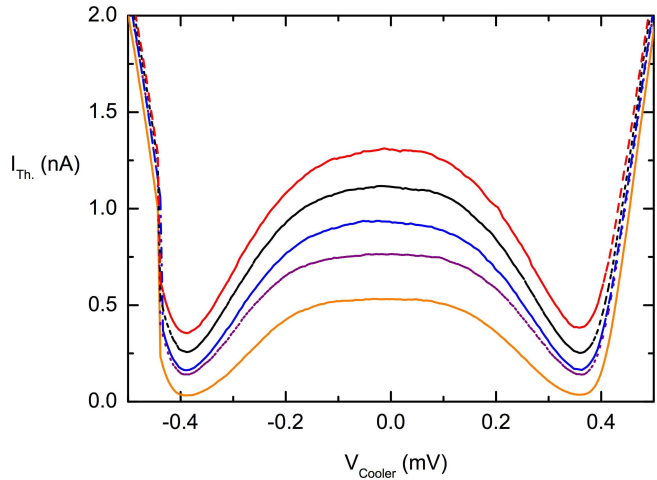}
\caption{(Color online) Sample A thermometer current versus cooler bias voltage for different base temperatures. From top to bottom, the cryostat temperature is 345, 325, 275, 306, 285 mK. The thermometer is voltage biased at 0.35 mV.}
\label{SampleA-cooling}
\end{figure}

Fig. \ref{SampleA-cooling} shows the measured thermometer current as a function of the cooler bias for cryostat temperatures from 275 to 345 mK, at a fixed thermometer voltage bias of 0.35 mV. The thermometer current decreases as the cooler bias increases towards its optimum bias, which indicates the cooling of the normal metal electrons. Further increase of the cooler bias injects hot quasi-particules in the normal metal, which heats it, as shown by the increase in the thermometer current.

S-I-N-I-S thermometers have been used in a variety of situations. Nahum et. al. \cite{APL-Nahum2} used it to show the first electronic cooling of a Cu island in a micro-cooler. Recently, N-I-S thermometer have been embedded in a LC resonant cicuit to achieve a sub-$\mu$s readout time \cite{APL-Schmidt}. The proximity effect near a S-N transparent junction is another possibility to measure the local electronic temperature \cite{APL-Jiang}. In the following, we propose a new concept to extract the electronic temperature in the context of S-I-N-I-S micro-coolers.

\section{Experimental results in S-I-N-I-S micro-coolers}

In the previous section, we used an additional double N-I-S junction as a thermometer to obtain the temperature of the cooled normal metal electrons. However, such a thermometer complicates the design and may inhibit a better understanding of the cooling behaviour. In this section, we will discuss the experiments done on a design with no external thermometer. As will be discussed afterwards, the electronic temperature will be obtained directly from the current voltage characteristic of the cooling junction.

\begin{figure}[t]
\centering
\includegraphics[width=0.29\linewidth]{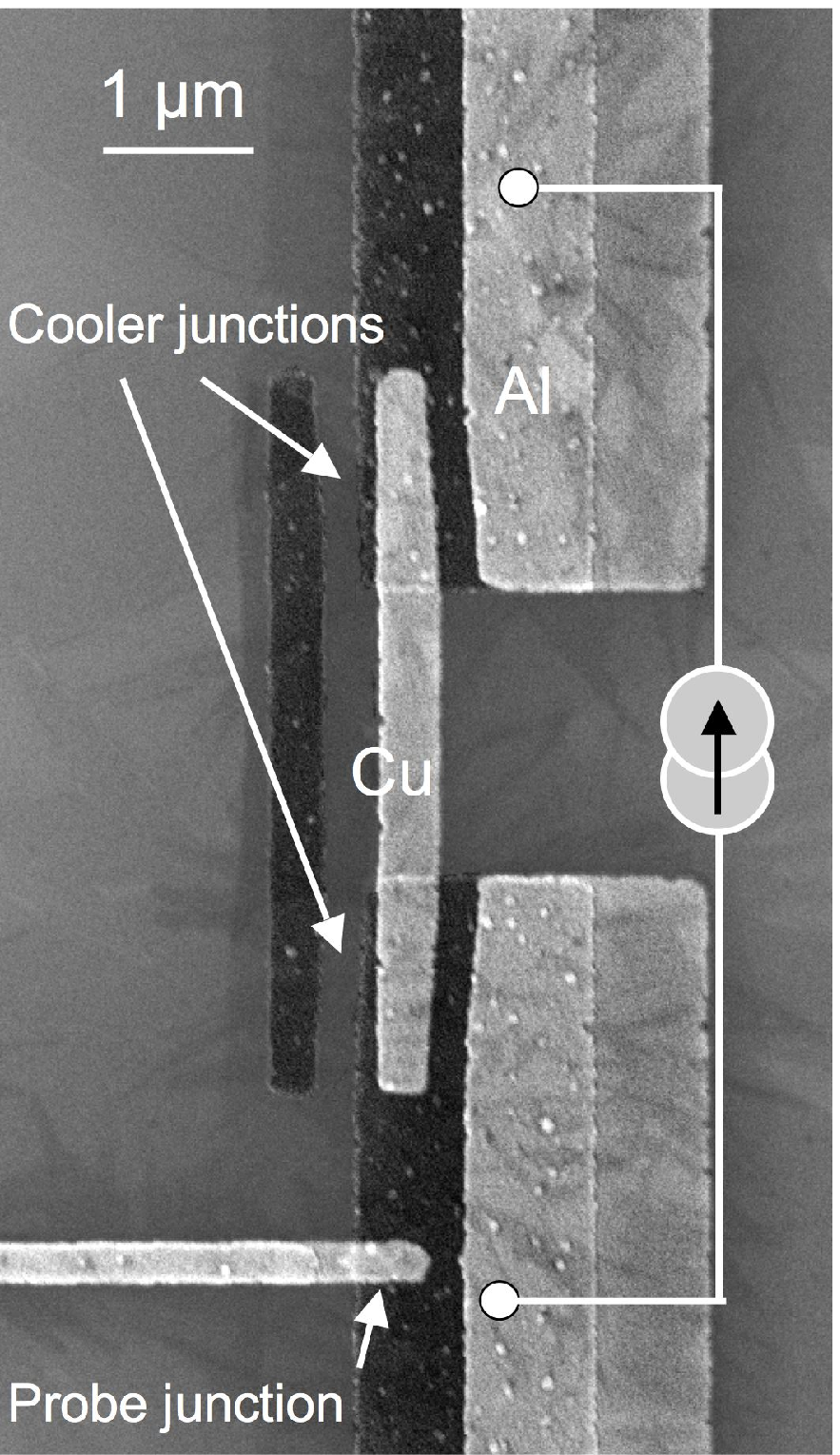}
\includegraphics[width=0.7\linewidth]{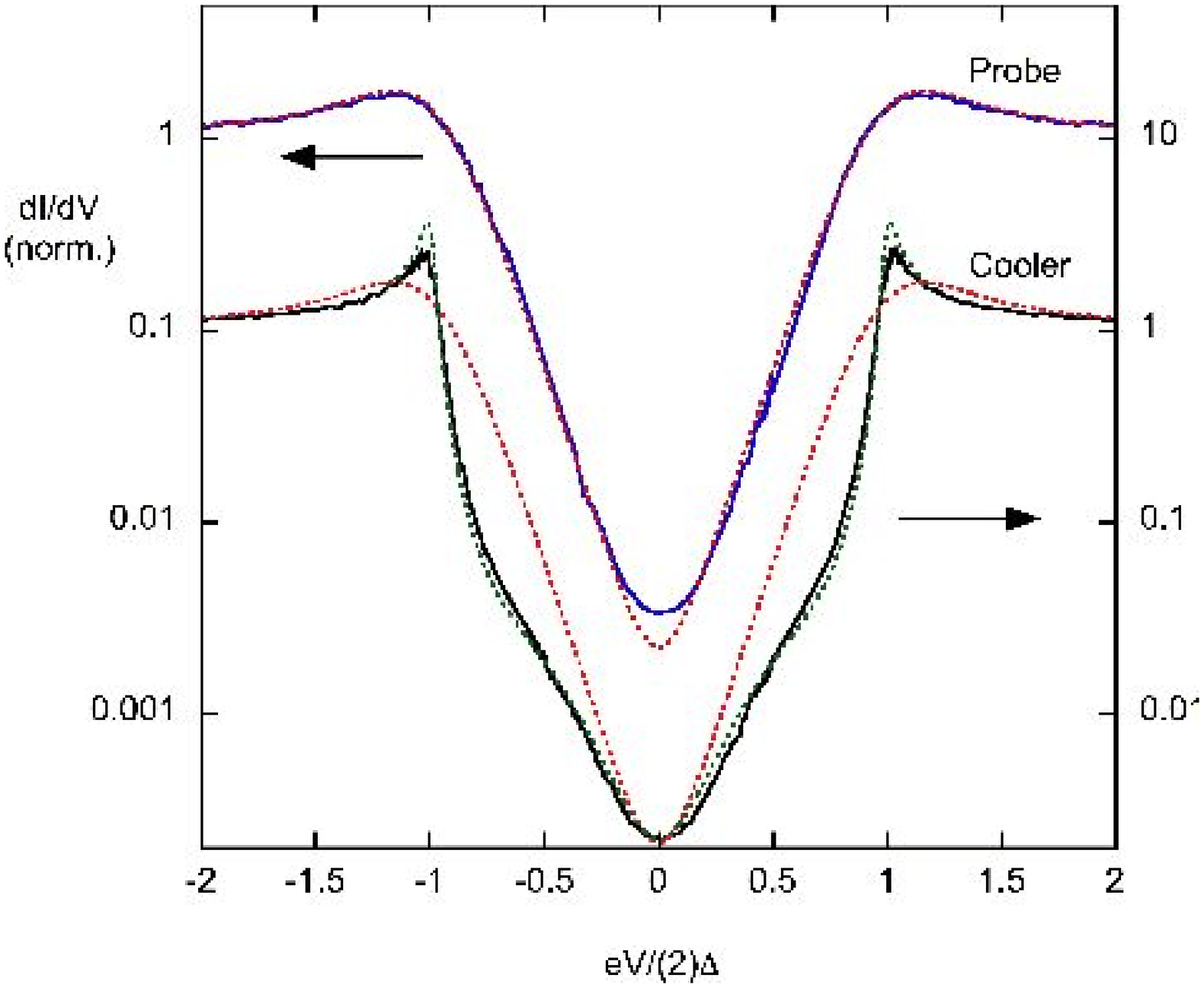}
\caption{(Color online) Left: Scanning electron micrograph of a cooler with no external thermometer on the central Cu island (Samples B and C). The cooler is made of two Al-AlO(x)-Cu junctions in series. The area of a cooler junction is 1.5 $\mu$m $\times$ 0.3 $\mu$m. In addition to the cooler, one of the three Al-AlO(x)-Cu probe junctions on the bottom superconducting electrode is visible. Right: Sample B differential conductance data obtained from the cooler and probe junctions at the cryostat temperature of 278 mK. Top curve: the full line shows the characteristic curve obtained from the probe junction and the dotted line is the calculated isotherm with R$_{n}$ = 5.7 k$\Omega$; $\Delta$ = 0.22 meV. Bottom curve: cooler junction data (full black line) compared with the calculated isotherm at T$_{e}$ = 300 mK with R$_{n}$ = 2.8 k$\Omega$; 2$\Delta$ = 0.42 meV (red dotted line). The green dotted nicely fitting the cooler characteristic is the fit with the thermal model in the device \cite{PRL-Sukumar1}. The voltage is normalized to $\Delta$ (top curve) and 2$\Delta$ (bottom curve).}
\label{fig:probendcooler}
\label{fig:sample_inh_thermo}
\end{figure}

Fig. \ref{fig:sample_inh_thermo} shows the scanning electron micrograph of one cooler device, where the central Cu island is attached to two superconducting reservoirs through tunnel junctions. The two 40 nm thick and 1.5 $\mu$m wide superconducting Al electrodes were in-situ oxidized in 0.2 mbar of oxygen for 3 min before the deposition of the central Cu island which is 5 (sample B) or 4 $\mu$m (sample C) long, 0.3 $\mu$m wide and 50 nm thick. In comparison with sample A, the Cu island is shorter while the cooler junctions area is the same, which leads to a more efficient surface/volume ratio in terms of electron cooling. In addition to these cooler junctions, we added three Cu tunnel probes of area 0.3 $\times$ 0.3 $\mu$m$^2$ on one Al electrode. Due to the large volume of the probe Cu island, the probe is strongly thermalized to the cryostat temperature. 

We have measured the current-voltage characteristic across either the cooler or the probe junction. The differential conductance for every junction is then numerically obtained. Fig. \ref{fig:probendcooler} shows in a logarithmic scale the differential conductance obtained from the cooler junction and one of the probe junction at a 278 mK cryostat temperature. The voltage axis is normalized to $\Delta$ (probe) or 2$\Delta$ (cooler) in order to nicely superpose the two curves. The probe junction characteristic shows the expected linear behavior for a quasi-equilibrium of the normal metal electrons. The data is fitted (dotted line) by an isotherm obtained from Eq. \ref{eq:symmetric-NIS} and at T$_{e}$ = 300 mK close to the cryostat temperature. Here an isotherm means that the electron distribution is at thermal equilibrium and thus given by Fermi distribution function.

In contrast, the isotherm does not fit the characteristic obtained from the cooler junction. The experimental data and the calculated isotherm coincide only at zero bias for sub gap bias. This is expected, since at zero bias there is no heat current due to tunneling, so that the electrons remain thermalized to the cryostat temperature. In the sub-gap region, the differential conductance of the cooler is smaller than the isotherm prediction. This demonstrates the cooling of the electronic population in the normal metal. 

\begin{figure}[t]
\centering
\includegraphics[width=0.7\linewidth]{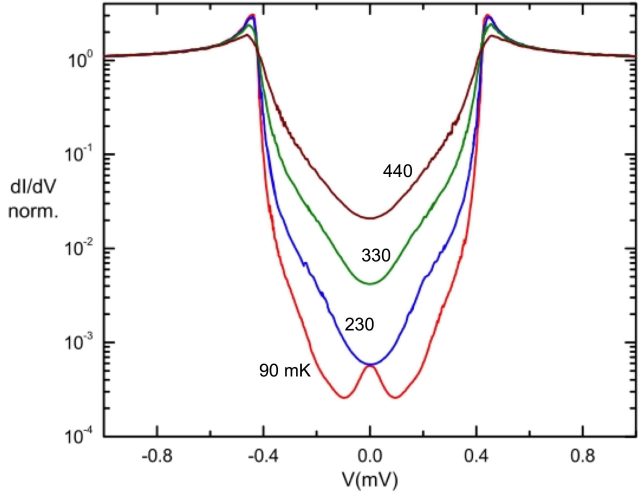}
\caption{(Color online) Sample C normalized differential conductance as a function of voltage bias of the cooler junction at the cryostat temperatures of 90, 230, 330 and 440 mK. The normal metal island is 4 $\mu$m long, 0.3 $\mu$m wide and 50 nm thick. The normal state resistance is $R_{N}$ = 1.9 k$\Omega$.}
\label{fig:dIdV-expt}
\end{figure}

Fig. \ref{fig:dIdV-expt} shows the differential conductance obtained from the cooler at different cryostat temperature, down to 90 mK \cite{PRL-Sukumar2}. Again, all the plots are on a logarithmic scale. As expected, the differential conductance at zero voltage bias decreases as the base temperature of the cryostat decreases. However, at very low temperature, the differential conductance at zero bias does not decrease further and even increases as temperature is decreased. This zero bias anomaly (see the T = 90 mK curve in Fig. \ref{fig:dIdV-expt}) cannot be explained by any ohmic leakage through the tunnel junctions, since such a leakage would give a constant background in the differential conductance plot. The same behavior was obtained in the probe junctions. 

\section{The electron energy distribution under tunneling-based cooling}

A normal metal cooled by electron tunneling is in principle in an out-of-equilibrium situation. The normal metal states are populated due to the electrons coming from the left superconducting electrode and depopulated by the right electrodes. The electron-electron inelastic scattering equilibrates the energy distribution towards a Fermi distribution. The electron-phonon coupling thermalizes the electron population to the phonon temperature of the metal. As a result, depending on the relative magnitudes of the injection, electron-electron scattering and electron-phonon scattering rates, different situations can be met.

If the electron-electron and electron-phonon scattering rates are small compared to the injection rate, then electrons occupy a non-equilibrium distribution $f_N(E)$ which can be very much different from the Fermi distribution. If the electron-electron or the electron-phonon scattering rate is large compared to the injection rate, the normal metal electrons follows a Fermi distribution $f_{o}(E,T_{e})$ at an electronic temperature $T_e$. In the latter case, two different regimes can be defined. If the electron-phonon relaxation time is short enough, the temperature of electrons in the normal metal T$_{e}$ is equal to the phonon temperature. This is the standard equilibrium situation. In the opposite case, electrons in the normal metal attain a temperature $T_e$ different from the phonon temperature. This is a quasi-equilibrium situation.

In quasi-equilibrium, if heat is injected in  the electron population, the electron temperature $T_e$ will be higher than the phonon temperature. A hot electron regime is then achieved \cite{PRL-Roukes,PRB-Wellstood}. In superconducting micro-coolers, heat is extracted from the normal metal electron population and a cold electrons regime is achieved. In every case, the temperature difference between electrons and phonons is of the order of $P/\Sigma U T^4$ where P is the power, $\Sigma$ is the electron-phonon constant, $U$ the metal volume.

In the following, we will solve the kinetic balance equation in the normal metal island to determine in a general way the electron energy distribution $f_N$. We will follow the description of Ref. \cite{PRB-Heslinga} to describe an electron distribution in the normal metal island of a S-I-N-I-S junction. We assume identical tunnel barriers and superconductors on either side of the superconductor. It is also assumed that the superconducting electrodes remain at thermal equilibrium so that the electron energy distribution is given by a Fermi-Dirac distribution f$_{o}$ at a temperature $T_S$.

The tunnel current from the superconductor to the normal metal (in a S-I-N-I-S junction) can be written as:
\begin{equation}
I(V) = \frac{2}{eR_{N}}\int^{\infty}_{-\infty} N_{S}(E-eV/2)[f_{0}(T_S,E - eV/2) - f_{N}(E)] dE.
\label{eq:non-NIS}
\end{equation}
Here $V$ is the voltage across, and $R_N$ is the resistance of, the two N-I-S junctions in series. The rate of population of a certain energy level due to injection from the superconductor to the normal metal is given by:
\begin{equation}
\frac{2}{R_{N}}N_{S}(E - eV/2)[f_{0}(T_S,E - eV/2) - f_{N}(E)],
\label{eq:inject-1}
\end{equation}
and the rate of depopulation of the energy level due to the extraction is given by:
\begin{equation}
\frac{2}{R_{N}}N_{S}(E + eV/2)[f_{N}(E) - f_{0}(T_S,E + eV/2)].
\label{eq:inject-2}
\end{equation}
In a relaxation time approximation, the inelastic relaxation of the injected quasiparticles at a certain energy level E is given by:
\begin{equation}
ALN_{N}(E)e^{2}\frac{f_{N}(E) - f_{0}(T_e,E)}{\tau_{E}},
\label{eq:inject-3}
\end{equation}
where A, L are the cross-section and the length of the normal metal, $N_N$ is the non-normalized density of electronic states, and $\tau_{E}$ the electron energy relaxation time. Here we assume that the electron energy relaxation in the normal metal can be considered as a coupling to a thermal bath at an effective temperature $T_e$. This temperature $T_e$ is thus the temperature that the electron population would reach if the energy relaxation is strong enough. For instance, in the case of a dominating electron-phonon coupling, $T_e$ would be equal to the phonon temperature.

At steady state, Eq. \ref{eq:inject-1} = Eq. \ref{eq:inject-2} + Eq. \ref{eq:inject-3}. Thus we get:
\begin{equation}
f_{N}(E) = \frac{N_{S}(E-\frac{eV}{2})f_{0}(T_S,E-\frac{eV}{2}) + N_{S}(E+\frac{eV}{2})f_{0}(T_S,E+\frac{eV}{2}) + \frac{f_{0}(T_e,E)}{\tau_{E}\Gamma}}{N_{S}(E-\frac{eV}{2}) + N_{S}(E+\frac{eV}{2}) + \frac{1}{\tau_{E}\Gamma}},
\label{eq:nonfermi}
\end{equation}
where the quantity $\Gamma$ given by:
\begin{equation}
\Gamma = \frac{2}{N_{N}(E_{F})R_{N}ALe^{2}}
\end{equation}
can be understood physically by noting that $\Gamma^{-1}$ is the mean residency time of an electron in the normal metal.

\begin{figure}[t]
\centering
\includegraphics[width=0.6\linewidth]{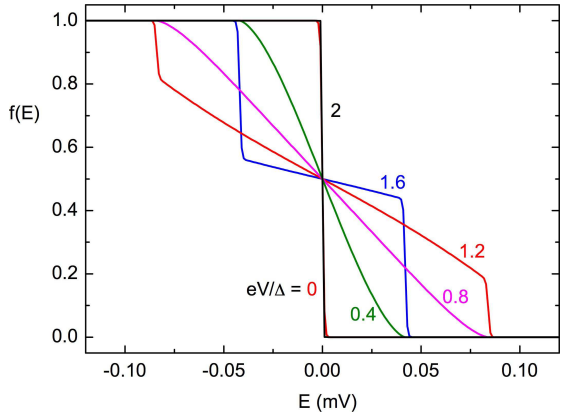}
\caption{(Color online) Calculated full non-equilibrium ($\tau_{E}\Gamma \rightarrow \infty$) electron energy distribution function in the normal metal for different bias eV/$\Delta$ = 0.0, 0.4, 0.8, 1.2, 1.6 and 2. Parameters for plot: $T_e$ = 0, $\Delta$ = 0.21 meV.}
\label{fig:nonfermi}
\end{figure}

Eq. \ref{eq:nonfermi} agrees with the one obtained by J. P. Pekola et. al in Ref. \cite{PRL-Pekola} for completely non-equilibrium distribution in N-metal. Eq. \ref{eq:nonfermi} describes a crossover from complete non-equilibrium to equilibrium distribution depending on the thermalization of the injected electrons in the central metallic island. If $\tau_{E}\Gamma <<$ 1, the normal metal electrons follow a Fermi distribution function. For $\tau_{E}\Gamma >>$ 1, the distribution function in the normal metal is different from the equilibrium Fermi distribution function. Fig. \ref{fig:nonfermi} displays the distribution function in this limit for different biases and at zero electronic temperature. The broadening of the distribution function up to eV/$\Delta$ = 1 and its sharpening for higher bias can be observed. 

\begin{figure}[t]
\centering
\includegraphics[width=0.495\linewidth]{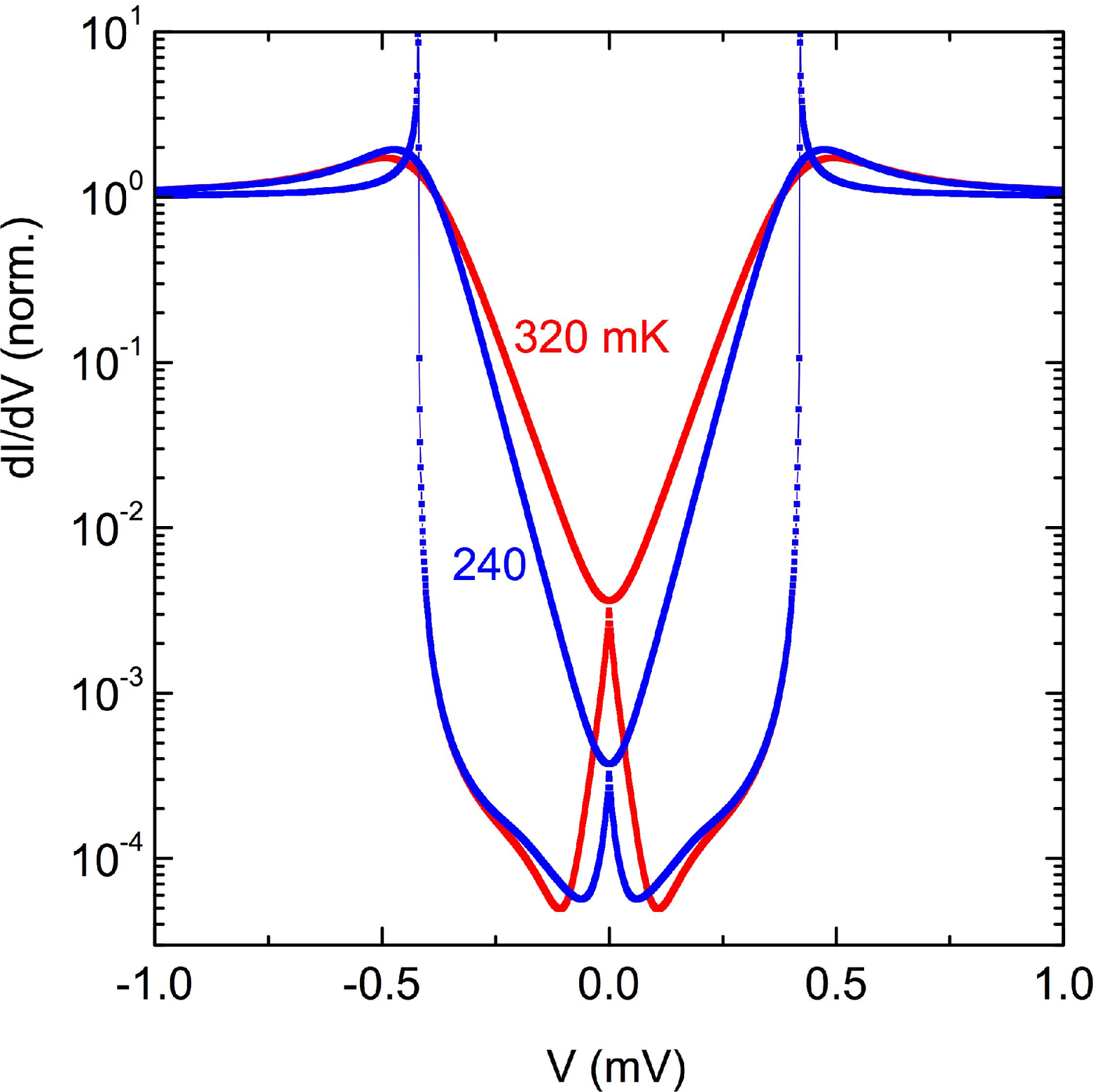}
\includegraphics[width=0.495\linewidth]{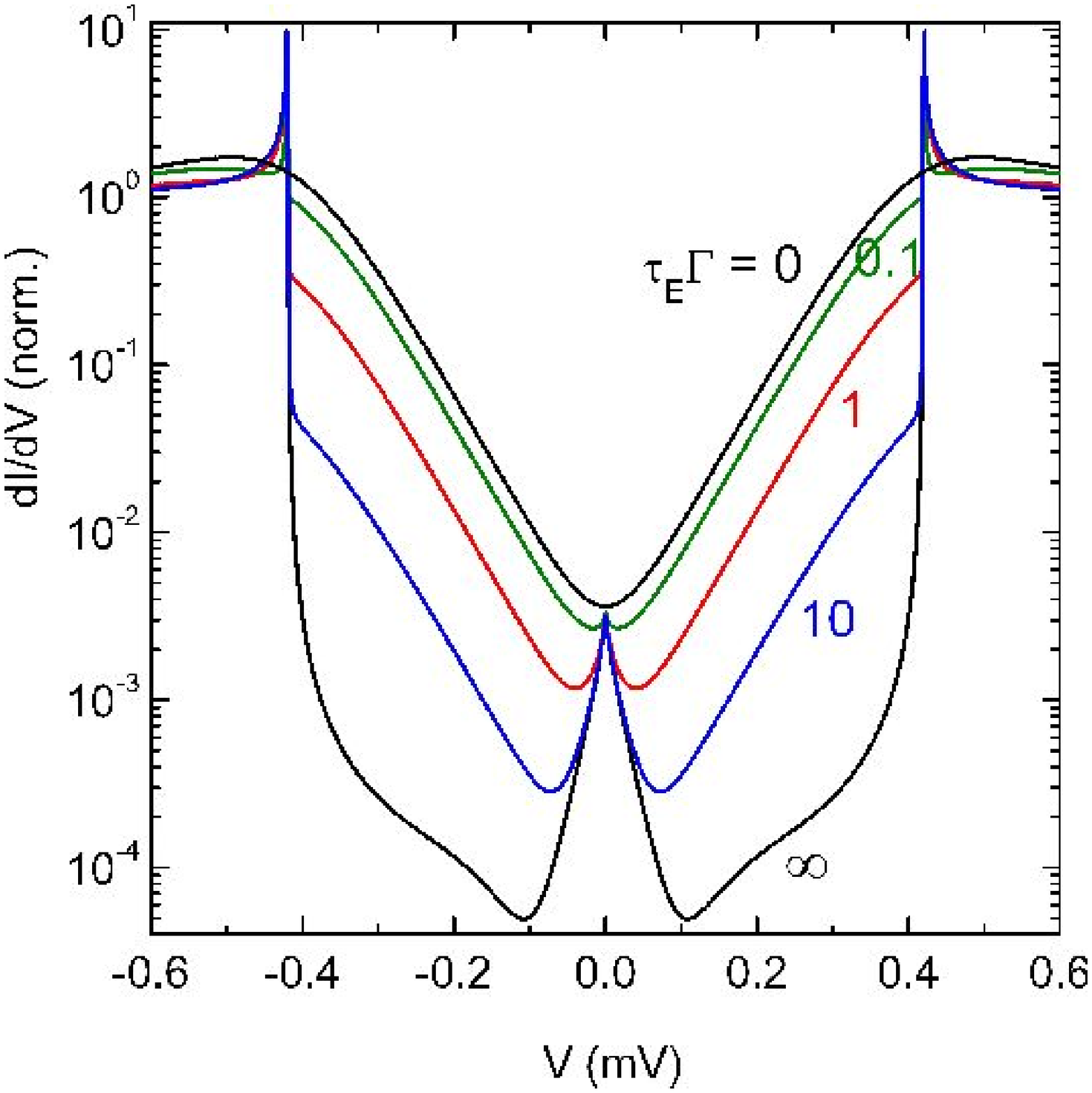}
\caption{(Color online) Left: Calculated differential conductance versus voltage bias plot for a complete non-equilibrium ($\tau_{E}\Gamma \rightarrow \infty$, lower curve at small bias) and a complete equilibrium ($\tau_{E}\Gamma \rightarrow 0$, upper curve at small bias) distribution function in the normal metal at the temperatures of 320 and 240 mK. At a given temperature, the two curves coincide at zero bias. The superconducting gap is $\Delta$ = 0.21 meV. Right: Calculated differential conductance versus voltage bias plot for different ratios of the relaxation rate to the injection rate in the normal metal. From bottom to top, the lines correspond to $\tau_{E}\Gamma$ = $\infty$ (black), 10 (blue), 1 (red), 0.1 (green) and 0 (completely thermalized case, black). The parameters are $\Delta$ = 0.21 meV, T$_{S}$ = 320 mK.}
\label{fig:dIdV_diffK}
\label{fig:fermi-nonfermi}
\end{figure}

The current through the N-I-S junction is obtained by substituting Eq. \ref{eq:nonfermi} in Eq. \ref{eq:non-NIS}. Fig. \ref{fig:fermi-nonfermi} left part displays the differential conductance versus the voltage bias for both complete equilibrium ($\tau_{E}\Gamma \rightarrow 0$) and full non-equilibrium ($\tau_{E}\Gamma \rightarrow \infty$) distribution in the normal metal at two different electronic bath temperatures. In the non-equilibrium case, there is a clear peak in the differential conductance near zero bias. Fig. \ref{fig:dIdV_diffK} right part shows the differential conductance isotherm for different $\tau_{E}\Gamma$ from 0 to $\infty$ at T = 320 mK. As the relaxation time for an electron decreases, the isotherm goes from completely non-equilibrium to quasi-equilibrium distribution in the central metallic island. In every case, the calculated non-equilibrium differential conductance is always below the thermal equilibrium regime one. For $\tau_{E}\Gamma <<$ 0.1, it is extremely difficult to distinguish the quasi-equilibrium curve from the fully isotherm curve.

A behavior similar to Fig. \ref{fig:dIdV_diffK} was observed in Ref. \cite{PRL-Pekola} experiments and discussed partly in terms of out-of-equilibrium effects. In the following, we will argue and present evidence, which shows that our experiments are not in this regime and that the electrons in the cooling metallic island are at quasi-equilibrium.

Comparing our experimental data to the theoretical predictions implies the determination of the physical parameters of our samples. The normalized density of states is given by $N_{N}(E_{F}) = (2mk_{F})/(2h^{2}\pi^{2})$. In sample B, we have R$_{N}$/2 = 1.4 k$\Omega$; k$_{F}$ = 1.75 $\times$ 10$^{10}$ m$^{-1}$; A.L = 5 $\times$ 0.3 $\mu$m$^2 \times$ 50 nm. We thus find $\Gamma$ to be of the order of 10$^{7}$ s$^{-1}$ in our samples. As $\Gamma$ is directly proportional to the the junction conductance, it would be much higher in the case of a S-N-S junction with transparent interfaces.

The inelastic time of scattering in the normal metal can be estimated separately by the weak localization correction to the resistance. We have performed magneto-resistance measurements on a copper wire made with the same source used for making our micro-coolers samples. The copper is of purity 6N (99.9999 percent). The phase coherence time $\tau_{\phi}$ is obtained by fitting the data using weak localisation theory. We find $\tau_{\phi}$ = 150 ps at 275 mK, which is consistent with expectations \cite{PRB-Pannetier,PRB-Pierre}. Taking into account the measured electron diffusion constant of 100 $cm^2/s$, we obtain a phase coherence length $L_{\varphi}$ of 1.5 $\mu m$. As we are interested here only in orders of magnitude, we identify the inelastic time to the phase coherence time, although the inelastic scattering time may be larger. In this way, we obtain a $\tau_{E}\Gamma$ parameter of the order of 0.015. As Fig. \ref{fig:dIdV_diffK} shows that the effect of non-equilibrium is very weak already with $\tau_{E}\Gamma$ = 0.1, we conclude that, in our case, the electrons in the normal metal can be considered as being at thermal equilibrium. 

\section {Inherent thermometry: discussion of the required assumptions}

In order to achieve a simple extraction of the electronic temperature from the current-voltage characteristic, we need to consider that Eq. \ref{eq:symmetric-NIS} gives the full description of the current. This implies the following assumptions: (i) the two refrigerating N-I-S junctions are symmetric so that there is an equal voltage drop across each junction; (ii) there are no leakage channels through the junction and no energy states within the superconducting gap; (iii) higher order tunneling processes can be neglected. In the following, we will discuss in detail these assumptions.

\begin{figure}[t]
\centering
\includegraphics[width=0.495\linewidth]{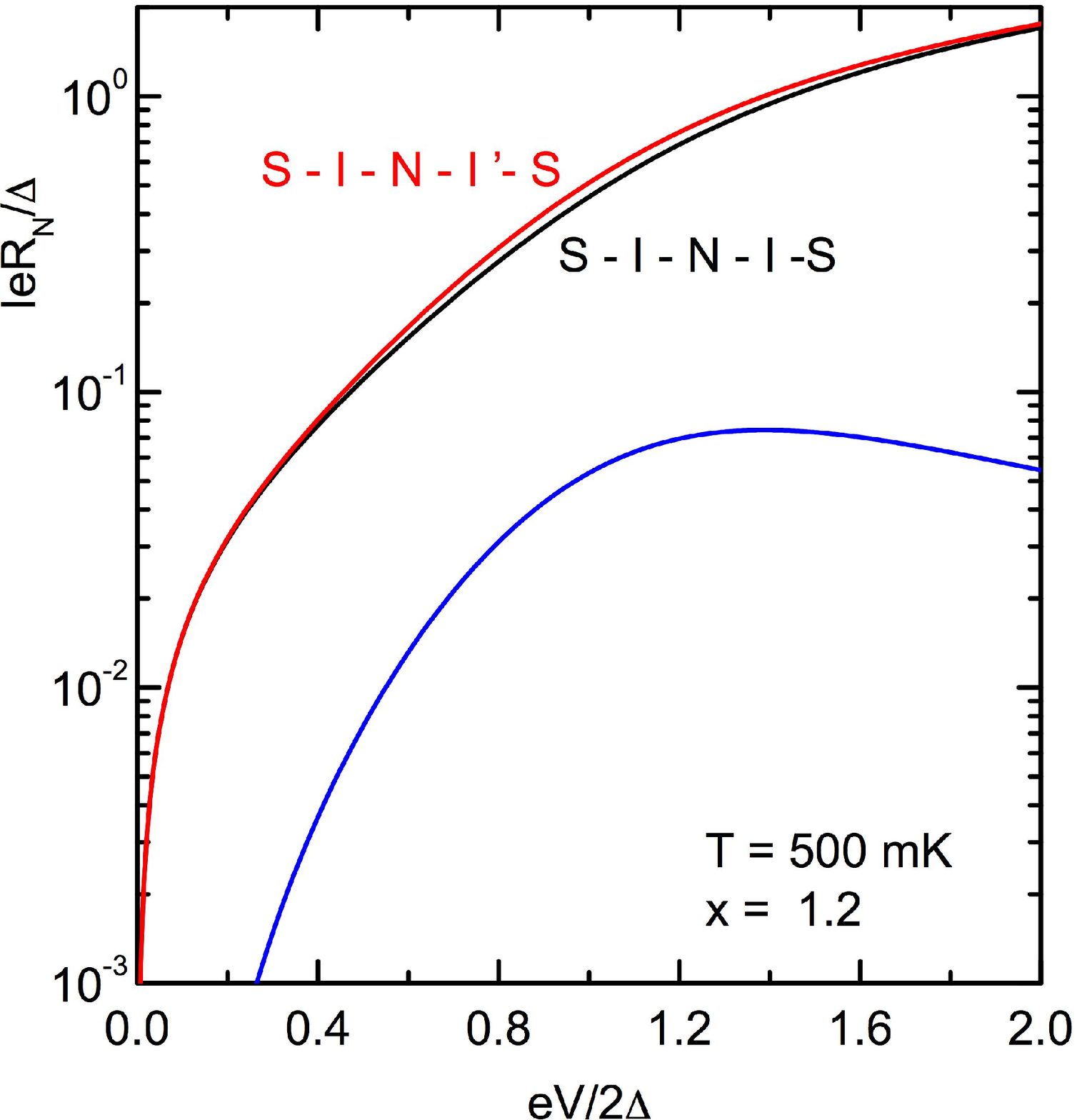}
\includegraphics[width=0.495\linewidth]{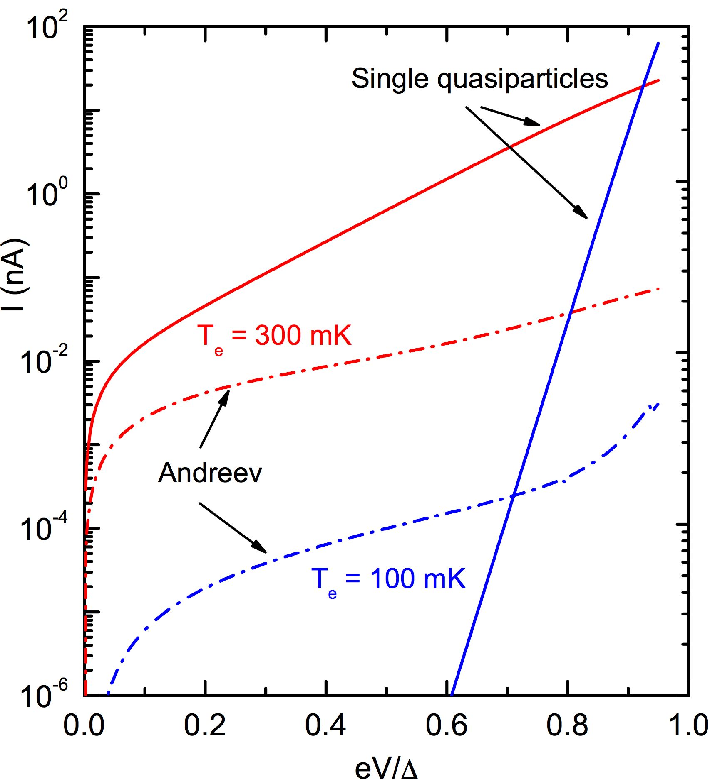}
\caption{(Color online) Left: The top two curves display the calculated normalized current - voltage characteristics of a non-symmetric S-I-N-I'-S (x = 1.2) and a symmetric S-I-N-I-S junction. The bottom curve displays the current difference as a function of voltage. Right: Calculated current contributions as a function of the voltage of a N-I-S junction at T$_{e}$ = 300 and 100 mK. The full lines show the contribution due to the single quasiparticle current and the dotted lines show the contribution of the phase-coherent Andreev current. The parameters are $\Delta$ = 0.225 meV, $R_n$ = 2 $k\Omega$ and $L_{\varphi}$ = 1.5 $\mu m$.}
\label{fig:2e-e-isotherm}
\label{fig:asymmetry}
\end{figure}

{\it Asymmetric tunnel junctions.} The central normal metal island is connected to superconducting reservoirs via two tunnel barriers. Although the two barriers are made at the same time, they can be slightly different from each other due for example to a different area. Let R$_{n1}$ and R$_{n2}$ be the normal state resistances of the two N-I-S junctions so that the total tunnel resistance across the device is R$_{n}$ = R$_{n1}$ + R$_{n2}$. The junction asymmetry is then defined by the ratio x = R$_{n1}$/R$_{n2}$. The total current across the S-I-N-I'-S junction is given by:
\begin{equation}
I(V) = \frac{1}{eR_{n}}\frac{1+x}{2x}\int^{\infty}_{0} N_{S}(E)[f_{N}(E - \frac{2x}{1+x}\frac{eV}{2}) - f_{N}(E + \frac{2x}{1+x}\frac{eV}{2})] dE.
\end{equation}
Fig. \ref{fig:asymmetry} shows the comparison of the calculated current-voltage characteristic for a symmetric and an asymmetric junction with x = 1.2. The total current across the junction for low bias voltage is little affected by the asymmetry. This is due to the singularity of the differential conductance at the gap, which regularizes the voltage \cite{Physica-Pekola}. The difference in the current is roughly proportional to the asymmetry and it is maximum at the gap voltage. For our samples, the resistance difference between the two junctions is estimated to be less than 5 percent. We consider then that we can neglect the effects of the device assymmetry.

{\it Energy states within the superconductor energy gap.} The presence of energy states within the superconductor gap can be described with the help of Dynes parameter $\eta$ so that \cite{PRL-Dynes}:
\begin{equation}
N_{S}(E) =| Re(\frac{E + i\eta}{\sqrt{(E + i\eta)^{2} + \Delta^{2}}})|.
\end{equation}

The effect of taking into account a significant Dynes parameter is to smear the differential conductance curve as a function of voltage and therefore increase the conductance at zero bias. In the case of bulk Al, there is no justification for taking a significant Dynes parameter. Fig. \ref{fig:dIdV-expt} shows that the differential conductance at intermediate temperature (above 250 mK) is definitely not influenced by a possible Dynes parameter since the differential conductance in this temperature range coincides with expectation and is also higher than the one observed at lower temperature (90 mK).

As already mentioned, a leakage in our tunnel junctions cannot explain the observed behavior with a differential conductance peak at zero bias. 

{\it Tunnel current due to higher order processes.} The tunneling of single-particles gives a zero tunnel current for energies below the superconducting gap $\Delta$: only electrons with energy E $> \Delta$ contribute to the current across the N-I-S junction. However, higher order tunnel processes do contribute to the current at an energy below the superconducting gap. The second order tunneling process  is the Andreev reflection \cite{Andreev,James}, which allows the transfer of two electrons with E $< \Delta$ from N to form a Cooper pair in S. The reverse process is also possible and corresponds to the creation of Andreev pairs \cite{Superlattices-Courtois} in the normal metal.

The amplitude of the Andreev reflection current vanishes with the transparency of the junction in the ballistic regime \cite{BTK}. However, confinement by the disorder in the electrode leads to the coherent addition of the many individual transmission probabilities for the transfer of a 2e charge, resulting in the enhancement of the sub-gap conductivity \cite{Physica-Volkov,PRL-vanWees,PRB-Hekking}. Fig. \ref{fig:2e-e-isotherm} right part shows the calculated current voltage characteristic of a N-I-S junction at two different temperatures and for a transparency of 10$^{-5}$ similar to the ones of our samples. Here we have used the theory of Ref. \cite{PRB-Hekking} in the 1D limit. The respective contributions of the single quasiparticle current and the phase coherent Andreev current are displayed. At moderate temperatures (see the 300 mK curve), the contribution due to Andreev current is found to be negligibly small in comparison with the single quasiparticle tunnel current. This does not hold at low enough temperatures, as is illustrated by the 100 mK curves of Fig. \ref{fig:2e-e-isotherm} right part  \cite{PRL-Sukumar2}. 

In conclusion for this section, we have examined the different assumptions required to consider that the current through our S-I-N-I-S micro-coolers is related solely to single-particle tunneling and thus follows Eq. \ref{eq:symmetric-NIS}. We conclude that they are correct in the intermediate temperature regime, which in our case corresponds to T $>$ 250 mK.

\section {Extraction of the electron temperature}

The above statement allows us to extract the temperature T$_{e}(V)$ in the sub-gap region by comparing our experimental current-voltage characteristic to a series of theoretical isotherm curves obtained from Eq. \ref{eq:symmetric-NIS}. 

\begin{figure}[t]
\centering
\includegraphics[width=0.69\linewidth]{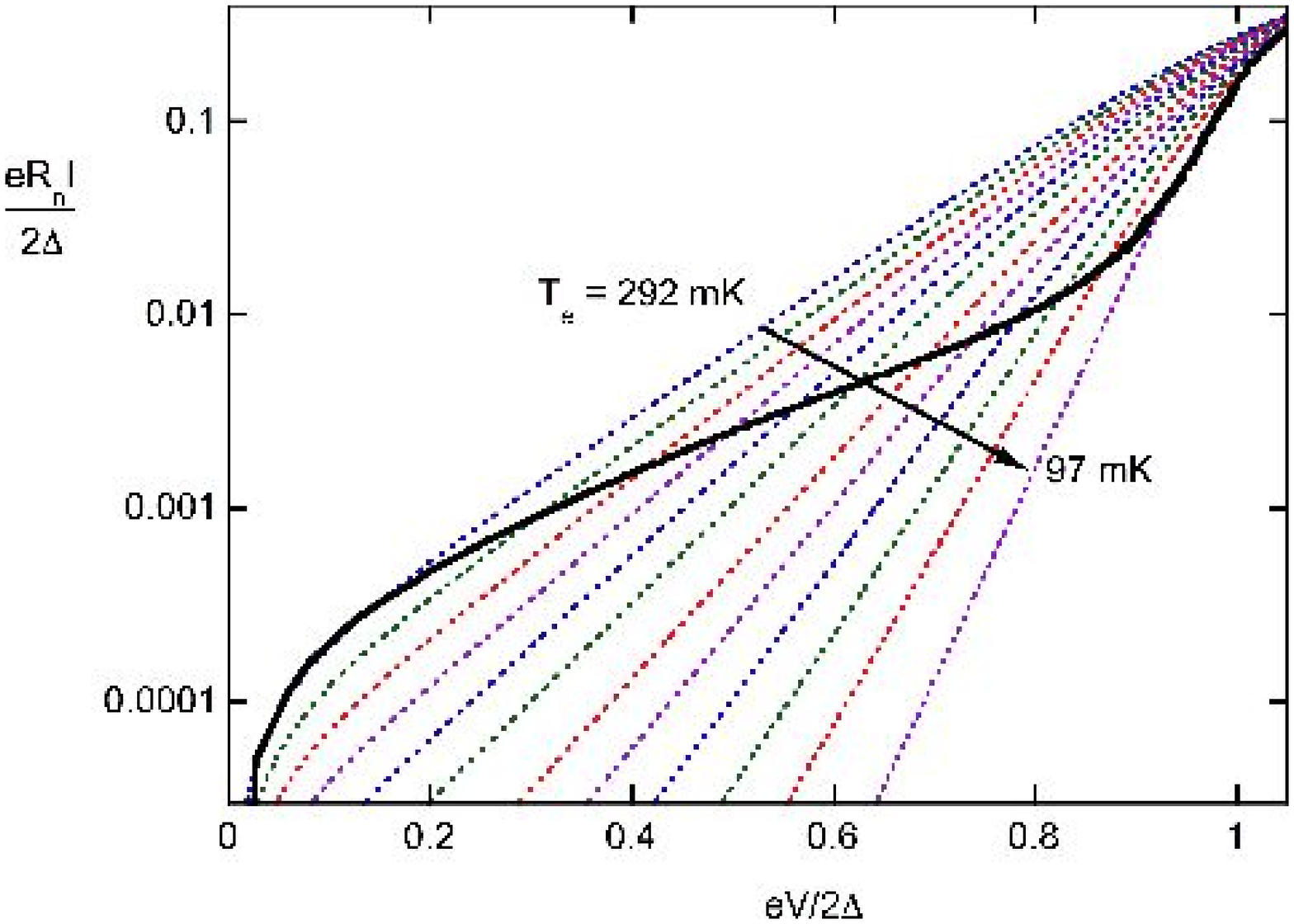}
\includegraphics[width=0.3\linewidth]{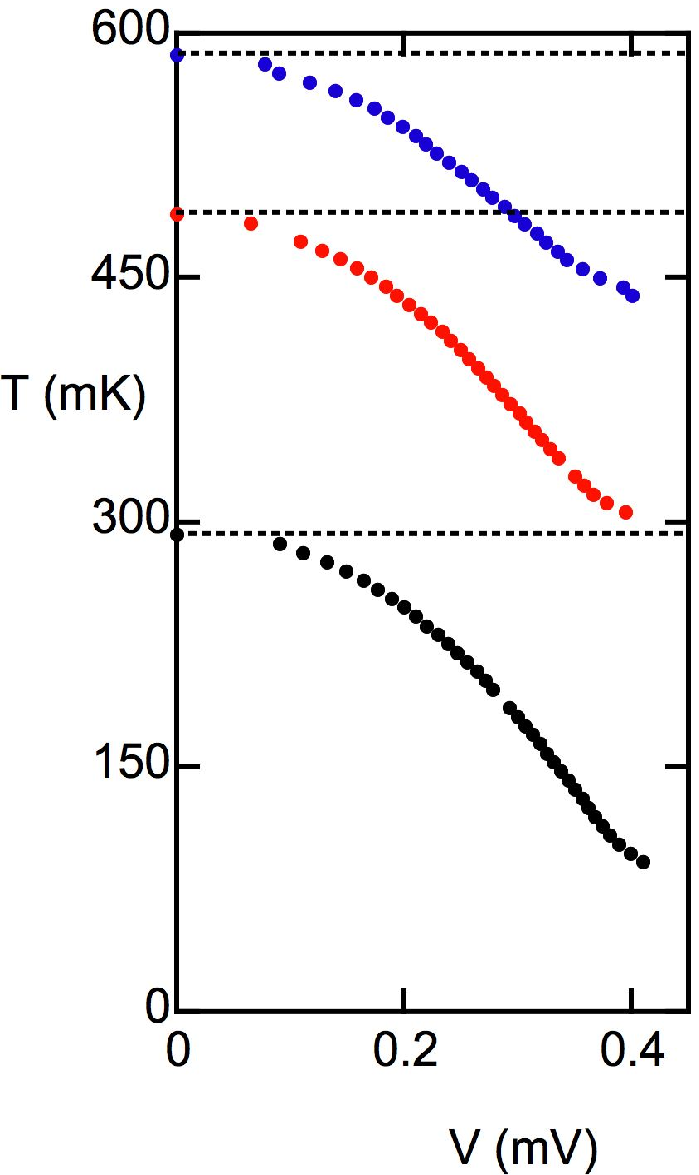}
\caption{(Color online) Left: Sample B experimental current-voltage characteristic at a 275 mK cryostat temperature (full line, black) superposed on a series of calculated isotherm characteristic following Eq. \ref{eq:symmetric-NIS} from T = 292 mK (top) to T = 97 mK (bottom). Every crossing point gives the electronic temperature T$_{e}$ in the central normal metal metal at a particular bias. Right: Sample B normal metal electronic temperature as a function of the cooler bias at a cryostat temperature of 275, 470 and 570 mK. The base temperature extracted from the data is 292, 489 and 586 mK respectively.}
\label{fig:expt_inherent}
\label{fig:expt_temp}
\end{figure}

Fig. \ref{fig:expt_inherent} left part shows the current voltage characteristic of the cooler junction at the cryostat temperature of 300 mK along with the simulated isotherms. At zero bias, there is a good overlap between the experiment (black complete line) and the isotherm corresponding a 292 mK electron temperature. As the cooler bias increases, the isotherm no longer follows the experiment and there is a crossover with the next isotherm at a slightly lower temperature, and so on. At the gap, electrons in the metal island have cooled from the base temperature of 292 mK to 97 mK. Here the normal state conductance for the S-I-N-I-S cooler of 2.8 k$\Omega$ and the gap 2$\Delta$ = 0.428 meV necessary for the isotherms calculations were obtained from the differential conductance plot.

It is interesting to note that such an extraction of the electron temperature cannot be done on a differential conductance-bias plot like Fig. \ref{fig:probendcooler}. As the tunneling current is a function of both bias and temperature, the full derivative gives:
\begin{equation}
\frac{dI}{dV} = \frac{\partial I(V,T)}{\partial V}+\frac{\partial I(V,T)}{\partial T}\frac{dT}{dV}
\end{equation}
Neglecting the second part on right side of the above equation would generate a systematic error in determining the electron temperature from the differential conductance.

Fig. \ref{fig:expt_temp} right part shows the Sample B electronic temperature as a function of the cooler bias voltage for three different cryostat temperatures. Every data point was obtained from the current voltage characteristic of the cooler junction. The black, red and blue symbols correspond to experiments at the cryostat temperatures of 275, 470 and 570 mK. The electronic temperature extracted from the data is, respectively, 292, 489 and 586 mK at zero bias, which is only slightly higher than the cryostat temperature. This experimental data set was further described in Ref. \cite{PRL-Sukumar1}.

\section{Conclusion}

We investigated the cooling of the central normal metal island in a S-I-N-I-S junction with \textbf{no external thermometer}. We have proposed a new technique to obtain the electron temperature in the normal metal by comparing the device current - voltage characteristic to the theoretical prediction. The electrons in the metal cool down from 300 mK to below 100 mK. This sample design with no external thermometer attains a higher ratio of volume/area of N-metal in comparison with previous design, which can contribute to a better cooling of the island.

\begin{acknowledgements}

We acknowledge the financial support from the ANR contract "Elec-EPR", the ULTI-3 and NanoSciERA "Nanofridge" EU projects. H. Courtois thanks the Low Temperature Laboratory for hospitality. We thank E. Favre-Nicollin and P.S. Luo for their contribution in the early stages of this work.

\end{acknowledgements}

\end{document}